\documentclass[epjCONF,columns]{svjour} 
\usepackage{graphics}
\usepackage[latin1]{inputenc}
\usepackage{savesym}
\usepackage{amsmath}
\savesymbol{iint}
\usepackage{txfonts}
\restoresymbol{TXF}{iint}

\session-title{Hadron Collider Physics Symposium 2011}
\begin{document}
\title{Measurement of the $\gamma$ angle from tree decays at the LHCb experiment}
\author{Alexandra Mart\'in S\'anchez\inst{1,2}\fnmsep\thanks{\email{sanchez@lal.in2p3.fr}} on behalf of the LHCb Collaboration}
%
\institute{Laboratoire de l'Acc\'el\'erateur Lin\'eaire, Orsay, France \and Universit\'e Paris-Sud 11, Orsay, France}
\abstract{
An overview of plans for the measurement of $\gamma$ at the LHCb experiment will be shown. The $\gamma$ angle is the least accurately known parameter of the CKM unitary triangle. The LHCb experiment at the CERN LHC aims to perform precision heavy flavour and \textit{CP} violation measurements, including improving the knowledge of~$\gamma$.  Focus will be put on methods using $B$ meson decays at the tree level, within the Standard Model framework. The early data recorded by the experiment, from \textit{pp} collisions at $\sqrt{s}$ = 7 TeV, allowed observations of the first signals of the $B$ decay modes that will be used to perform this measurement.
} 
%
\maketitle
\section{Introduction: the CKM matrix and the $\gamma$~angle}
\label{intro}
It is well established in particle physics that in the quark sector, the interaction eigenstates are not equal to the mass eigenstates, but a linear combination of them; this effect is known as \textit{quark mixing}. The matrix that allows to make the change of basis between interaction and mass eigenstates is the CKM matrix, \textit{c.f.} \cite{CKM paper}. It is a three by three unitary matrix 
\begin{equation}
\begin{pmatrix}
d' \\ s' \\ b'
\end{pmatrix}
= V_{CKM}
\begin{pmatrix}
d \\ s \\ b
\end{pmatrix}
=
\begin{pmatrix}
V_{ud} & V_{us} & V_{ub} \\ 
V_{cd} & V_{cs} & V_{cb} \\ 
V_{td} & V_{ts} & V_{tb}  
\end{pmatrix}
\begin{pmatrix}
d \\ s \\ b
\end{pmatrix}
\end{equation}
that can be parametrised by three real amplitudes and a phase; the Wolfenstein parametrization does this in a natural way
\begin{equation}
V_{CKM} =
\begin{pmatrix}
1-\lambda^2/2 & \lambda & A\lambda^3(\rho-i\eta) \\ 
-\lambda & 1-\lambda^2/2 & A\lambda^2 \\ 
A\lambda^3(1-\rho-i\eta) & -A\lambda^2 & 1  
\end{pmatrix}
+ \mathcal{O}(\lambda^4) ~
\end{equation}
as it takes into account the physical hierarchy of the different matrix elements, \textit{c.f.} \cite{Wolfenstein}.


The unitarity of this matrix demands
\begin{eqnarray}
\sum_i V_{ij}V_{ik}^*&=&\delta_{jk} ~~~\mathrm{and} \\
\sum_i V_{ji}V_{ki}^*&=&\delta_{jk} ~.
\end{eqnarray}

\begin{figure}
\begin{center}
\resizebox{0.75\columnwidth}{!}{%
  \includegraphics{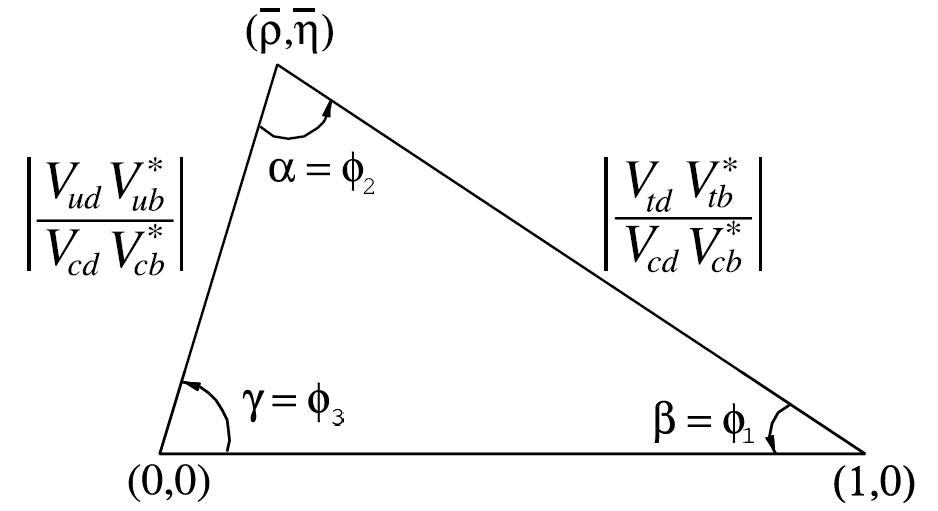}}
\end{center}
\caption{The CKM unitary triangle.}
\label{triangle}       
\end{figure}

\begin{figure}
\resizebox{1.00\columnwidth}{!}{%
  \includegraphics{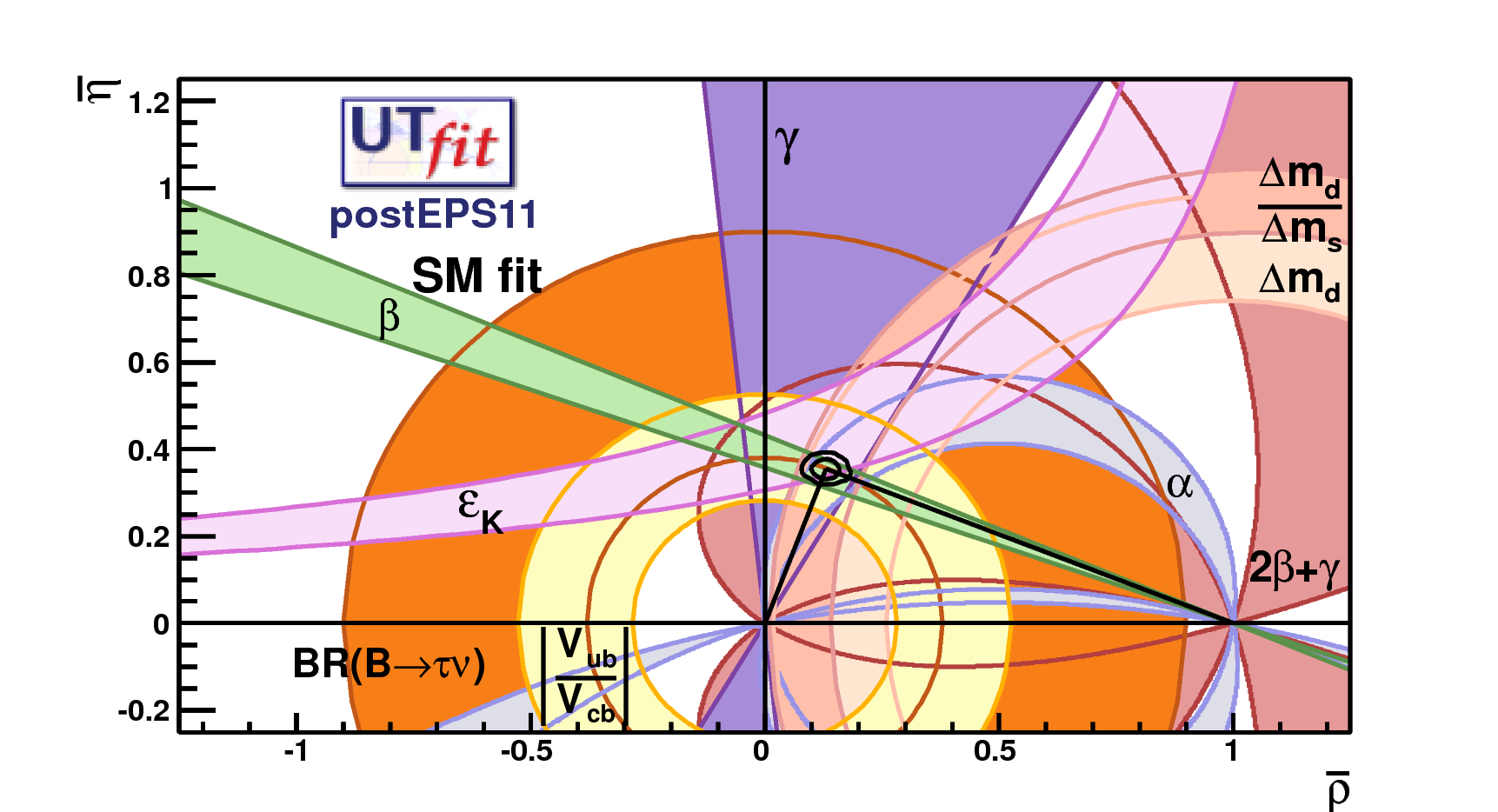}}
\caption{Status of the CKM unitary triangle.}
\label{utfit}       
\end{figure}

The six relations where $j \neq k$ can be represented as \textit{unitary triangles} in the complex plane. One in particular, corresponding to
\begin{equation}
V_{ud}V_{ub}^*+V_{cd}V_{cb}^*+V_{td}V_{tb}^*=0 ~\mathrm{,}
\end{equation}
is easier to study than the others, as the sides of the triangle are all of the same order $\lambda^3$, see figure \ref{triangle}.
The experiemental status of this unitary triangle is summarized in figure \ref{utfit}, \textit{c.f.}~\cite{UTfit} (see also \cite{CKMfitter}).

Regarding the $\gamma$ angle, the current value from global fits is
\begin{equation}
\gamma_{\mathrm{global~fit}} = (69 \pm 3)^\circ ~\mathrm{,}
\end{equation}
while the value coming from direct measurements is
\begin{equation}
\gamma_{\mathrm{direct~measurements}} = (76 \pm 11)^\circ ~.
\end{equation}
The uncertainty on the direct measurements of the $\gamma$ angle is still large compared to the other parameters of the CKM theory. One of the main goals of the LHCb experiment is to perform a precise direct measurement of the $\gamma$ angle. This will allow to further constrain the unitary triangle; inconsistencies between the different measurements can be indications of New Physics.

\section{The $\gamma$ measurement from trees at the LHCb experiment}
\label{sec:2}
The LHCb experiment is a one arm spectrometer designed to perform precision measurements of the Standard Model parametres specially on the beauty and charm sectors, \textit{c.f.} \cite{LHCb tdr}. It is one of the four main experiments on the Large Hadron Collider (LHC) at CERN in Geneva (Switzerland).

LHCb will be able to improve precision on the $\gamma$ measurement. This document will focus on measurements from tree diagrams, which are not sensitive to New Physics contributions (which is not the case in the measurements from diagrams with loops). The measurements can be performed by time-dependent or time-integrated approaches, and they consist on the study of the interference between $b \rightarrow u$ and $b \rightarrow c$ transitions giving the same final state (\textit{i.e.} through the $V_{ub}$ and $V_{cb}$ elements of the CKM matrix).

With the first data recorded at $\sqrt{s}$ = 7 TeV in 2010 and 2011 at LHCb several key decay modes have already been observed and some early measurements have been 
performed.

\subsection{Time-dependent measurements}
\label{sec:2.1}
\begin{figure}
\resizebox{1.00\columnwidth}{!}{%
  \includegraphics{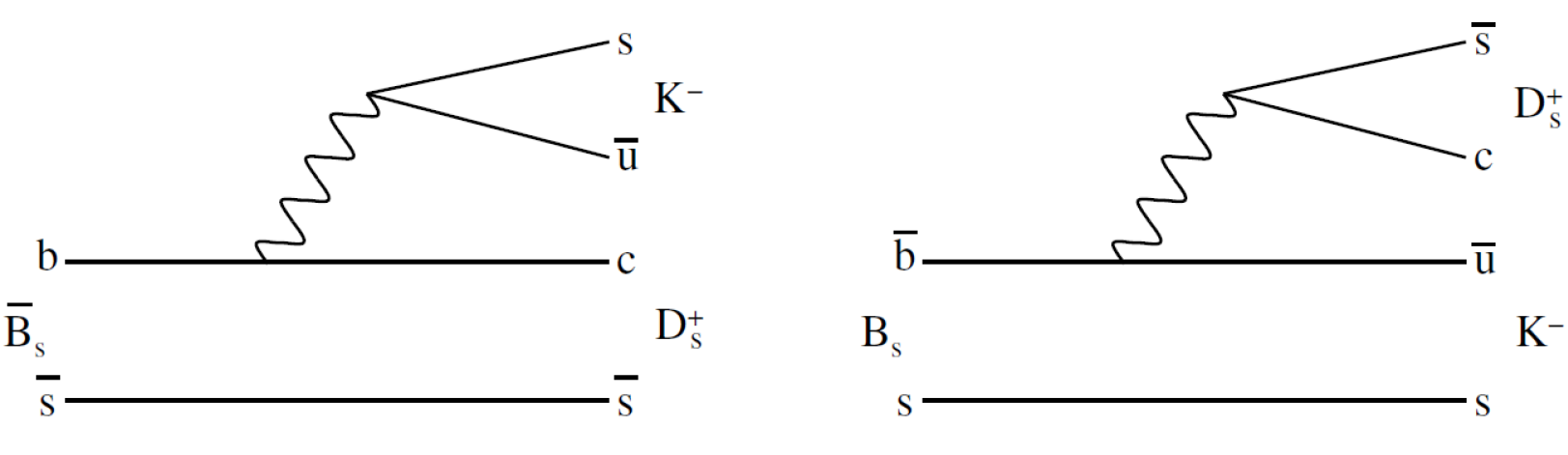}}
\caption{Feynman diagrams for $B_s^0 \rightarrow D_s K$ the decay.}
\label{timedep_diagram}       
\end{figure}

The interference between the direct decay and the decay after mixing (both accessing the same final state) is studied. The modes that give the best sensitivity to $\gamma$ and that are looked at by LHCb in this way are $B_s^0 \rightarrow D_s K$ and $B^0 \rightarrow D \pi$, see figure \ref{timedep_diagram}. 

\begin{figure}
\resizebox{1.00\columnwidth}{!}{%
  \includegraphics{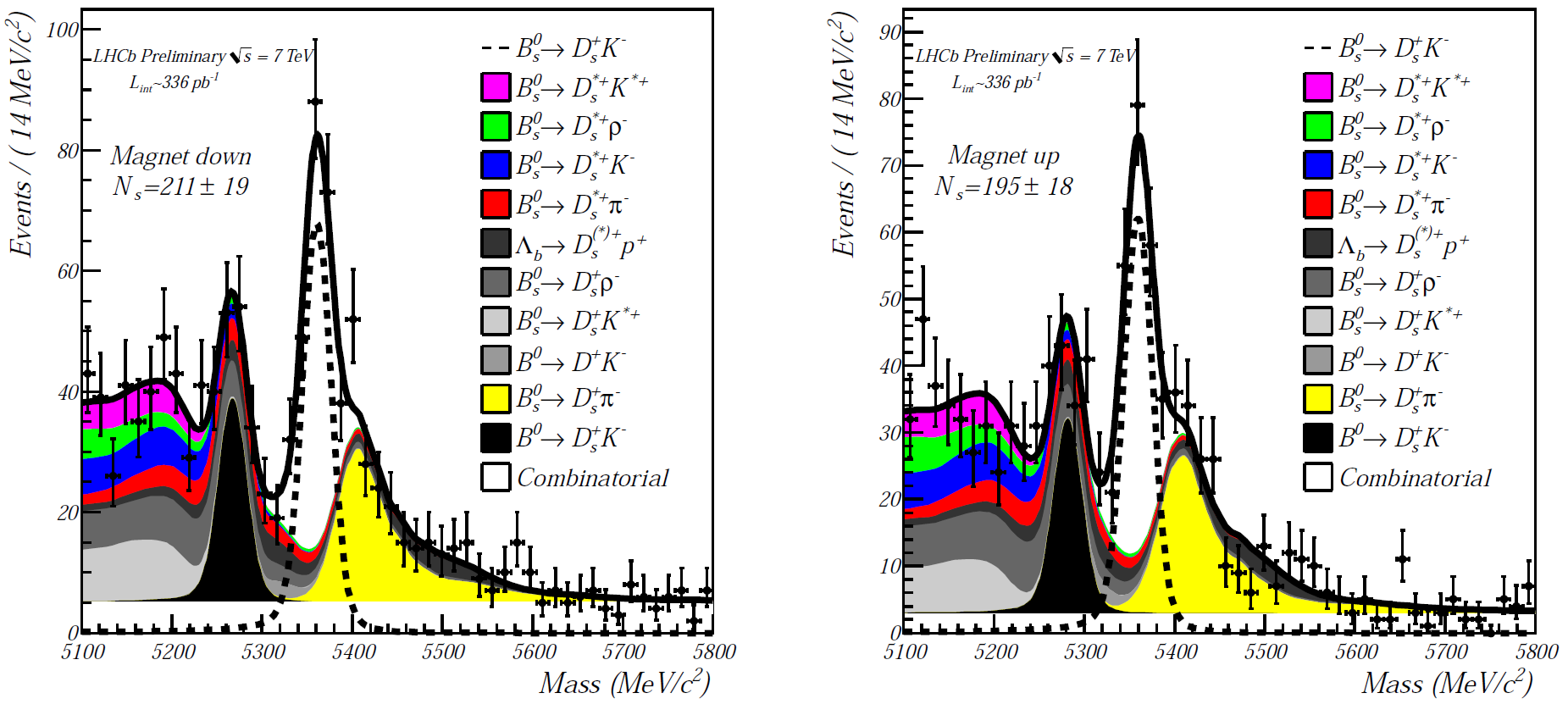}}
\caption{$B_s^0 \rightarrow D_s K$, $D_s^+ \rightarrow K^+ K^- \pi^+$ invariant mass distribution.}
\label{timedep_plot}       
\end{figure}

The branching ratio measurement of $B_s^0 \rightarrow D_s^{\mp} K^{\pm}$ has been performed with 336 $\mathrm{pb^{-1}}$ of data recorded at LHCb in 2011, in the $D_s^+ \rightarrow K^+ K^- \pi^+$ mode, obtaining 
\begin{eqnarray}
\mathcal{B}(B_s^0 \rightarrow D_s^{\mp} K^{\pm}) =  (1.97 \pm 0.18 (\mathrm{stat.})~^{+0.19} _{-0.20} (\mathrm{syst.})\\
 ^{+0.11} _{-0.10} (\textit{$f_s/f_d$})) \times 10^{-4} ~\mathrm{,} \nonumber
\end{eqnarray}
where the first uncertainty is statistical, the second reflects experimental systematics and the third is due to the uncertainty in the knowledge of the hadronisation fractions ($f_s/f_d$), see figure \ref{timedep_plot}, \textit{c.f.} \cite{timedep conf note}. Efforts will be put in the future on the time-dependant analysis of this decay to extract the measurement of $\gamma$.

\subsection{Time-integrated measurements}
\label{sec:2.2}
\begin{figure}
\resizebox{1.00\columnwidth}{!}{%
  \includegraphics{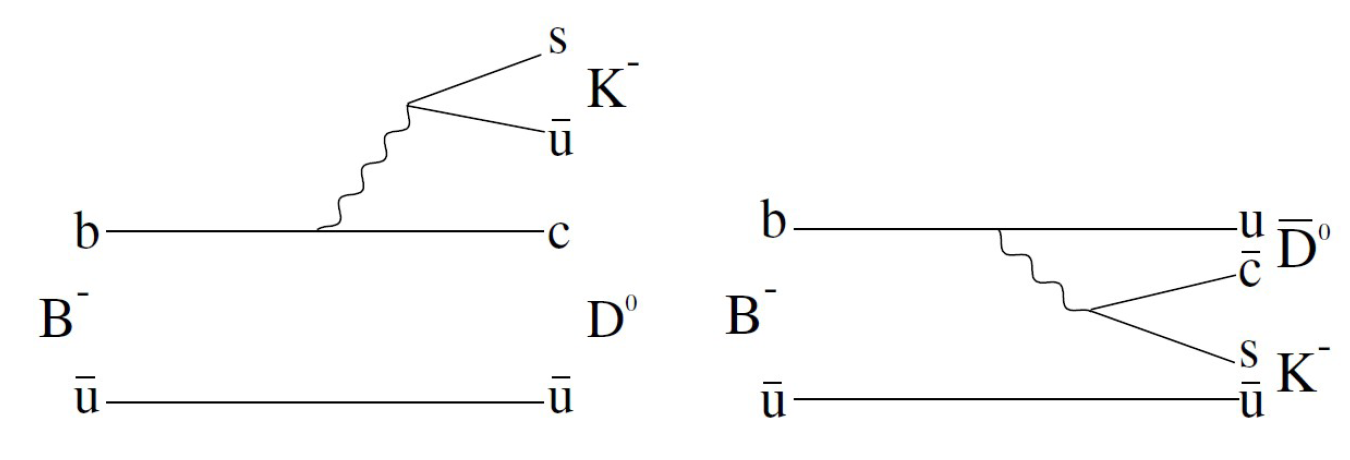}}
\caption{Feynman diagrams for $B^- \rightarrow D^0 K^-$ the decay.}
\label{timeintcharged_diagram}       
\end{figure}

\begin{figure}
\resizebox{1.00\columnwidth}{!}{%
  \includegraphics{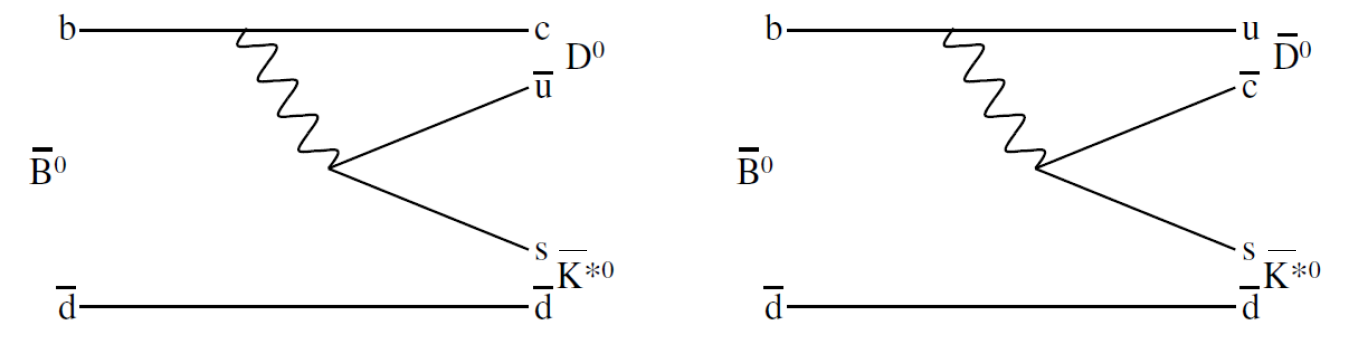}}
\caption{Feynman diagrams for $\overline{B}^0 \rightarrow D^0 \overline{K}^{*0}$ the decay.}
\label{timeintneutral_diagram}       
\end{figure}

Self-tagging modes are used, where the interference between decays to the same final products by different intermediate states gives access to $\gamma$. The decays $B^- \rightarrow \overline{D}^0 K^-$ and $\overline{B}^0 \rightarrow D^0 \overline{K}^{*0}$ are studied, see figures \ref{timeintcharged_diagram} and \ref{timeintneutral_diagram}: the first one has larger branching ratios but the interference between the two diagrams is smaller due to the fact that one of them is colour suppressed and the other one is not (so they are of different orders of magnitude); the second one is more difficult to study because the branching ratios are smaller, but the interference is larger as both diagrams are colour suppressed (and so they are of the same order).

There are different measurement methods depending on the decay mode of the $D^0$ meson, that will be explained in the following sections. In any case, the observables that give sensitivity to $\gamma$ are the asymmetries between the decay and its charged conjugate, $A$, and the ratio of the sum, $R$.

\subsubsection{$\gamma$ by the GLW method}
\label{sec:2.2.1}
The $D^0$ meson decays to \textit{CP} eigenstates: $D^0 \rightarrow K^+ K^-$, $D^0 \rightarrow \pi^+ \pi^-$, \textit{c.f.} \cite{GLW paper1},\cite{GLW paper2}. The asymmetry and ratio observables are given by
\begin{eqnarray}
A_{CP+} &=& \frac{\mathrm{\Gamma}(B^-\rightarrow D_\pm K^-) - \mathrm{\Gamma}(B^+\rightarrow D_\pm K^+)}{\mathrm{\Gamma}(B^-\rightarrow D_\pm K^-) + \mathrm{\Gamma}(B^+\rightarrow D_\pm K^+)} \nonumber \\
&=& \frac{\pm 2 r_B\sin\delta_B\sin\gamma}{1+r_B^2\pm 2 r_B\cos\delta_B\cos\gamma} ~~~\mathrm{and}
\end{eqnarray}
\begin{eqnarray}
R_{CP+} &=& 2\frac{\mathrm{\Gamma}(B^-\rightarrow D_\pm K^-) + \mathrm{\Gamma}(B^+\rightarrow D_\pm K^+)}{\mathrm{\Gamma}(B^-\rightarrow D^0 K^-) + \mathrm{\Gamma}(B^+\rightarrow D^0 K^+)} \nonumber \\
&=& 1 + r_B^2 \pm 2 r_B\cos\delta_B\cos\gamma ~.
\end{eqnarray}

\begin{figure}
\resizebox{1.00\columnwidth}{!}{%
  \includegraphics{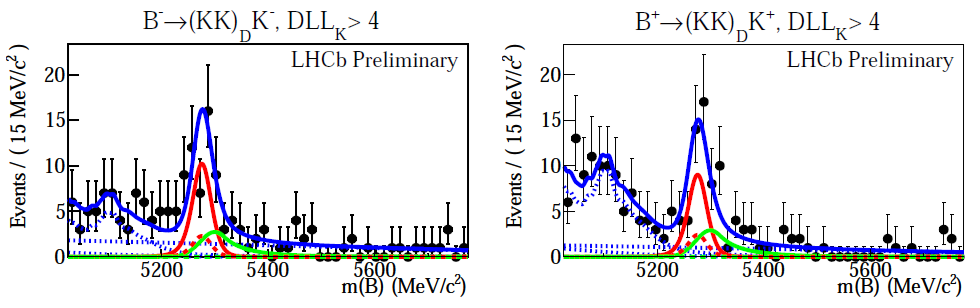}}
\caption{$B \rightarrow D^0 K$, $D^0 \rightarrow K^+ K^-$ invariant mass distribution.}
\label{GLWplot}       
\end{figure}

The GLW $D^0 \rightarrow K^+ K^-$ mode has been measured at LHCb with 36.5 $\mathrm{pb^{-1}}$ data recorded in 2010. The preliminary results are
\begin{eqnarray}
A_{CP+} &=& 0.07 \pm 0.18(\mathrm{stat.}) \pm 0.07(\mathrm{syst.}) ~\mathrm{,}\\
R_{CP+} &=& 1.48 \pm 0.31(\mathrm{stat.}) \pm 0.12(\mathrm{syst.}) ~\mathrm{,}
\end{eqnarray}
where the first uncertainty is statistical and the second systematic, see figure \ref{GLWplot}, \textit{c.f.} \cite{GLW conf note}. The asymmetry result is consistent with zero.

\subsubsection{$\gamma$ by the ADS method}
\label{sec:2.2.2}
The $D^0$ decays to flavour specific final states are considered: $D^0 \rightarrow K \pi$, \textit{c.f.} \cite{ADS paper}. Here the asymmetry and ratio observables are 
\begin{eqnarray}
A_{ADS} &=& \frac{\mathrm{\Gamma}(B^-\rightarrow D(K^+\pi^-) K^-) - \mathrm{\Gamma}(B^+\rightarrow D(K^-\pi^+) K^+)}{\mathrm{\Gamma}(B^-\rightarrow D(K^+\pi^-) K^-) + \mathrm{\Gamma}(B^+\rightarrow D(K^-\pi^+) K^+)} \nonumber \\
&=& \frac{2 r_B r_D \sin(\delta_B+\delta_{K\pi})\sin\gamma}{r_B^2 + r_D^2 + 2 r_B r_D cos(\delta_B+\delta_{K\pi})\cos\gamma} ~~~\mathrm{and}
\end{eqnarray}
\begin{eqnarray}
R_{ADS} &=& \frac{\mathrm{\Gamma}(B^-\rightarrow D(K^+\pi^-) K^-) + \mathrm{\Gamma}(B^+\rightarrow D(K^-\pi^+) K^+)}{\mathrm{\Gamma}(B^-\rightarrow D(K^-\pi^+) K^-) + \mathrm{\Gamma}(B^+\rightarrow D(K^+\pi^-) K^+)} \nonumber \\
&=& r_B^2 + r_D^2 + 2 r_B r_D \cos(\delta_B+\delta_{K\pi})\cos\gamma ~.
\end{eqnarray}

\begin{figure}
\resizebox{1.00\columnwidth}{!}{%
  \includegraphics{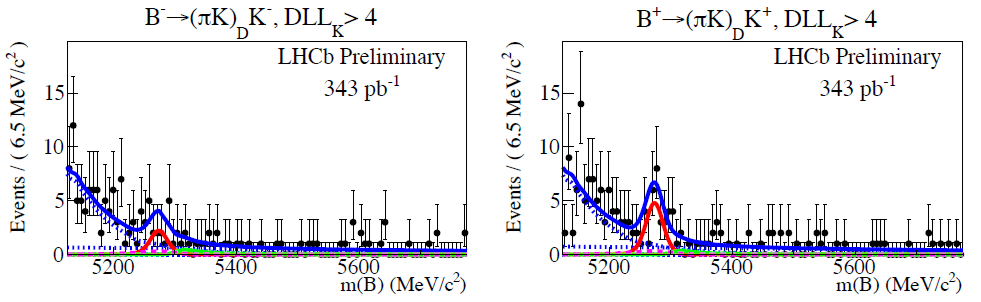}}
\caption{$B \rightarrow D^0 K$, $D^0 \rightarrow K^- \pi^+$ invariant mass distribution.}
\label{ADSplot}       
\end{figure}

The preliminary results of LHCb on these quantities with 343 $\mathrm{pb^{-1}}$ of data recorded in 2011 are
\begin{eqnarray}
A_{ADS} &=& -0.39 \pm 0.17(\mathrm{stat.}) \pm 0.02syst.) ~\mathrm{,}\\
R_{ADS} &=& (1.66 \pm 0.39(\mathrm{stat.}) \pm 0.24(\mathrm{syst.})) \times 10^{-2} ~\mathrm{,}
\end{eqnarray}
see figure \ref{ADSplot}, \textit{c.f.} \cite{ADS conf note}. 

The ADS method can also be applied to other $D^0$ modes such as $D^0 \rightarrow K^- \pi^+ \pi^- \pi^+$, that is also being investigated at LHCb.

\subsubsection{The neutral mode $\overline{B}^0 \rightarrow D^0 \overline{K}^{*0}$}
\label{sec:2.2.3}
\begin{figure}
\begin{center}
\resizebox{0.75\columnwidth}{!}{%
  \includegraphics{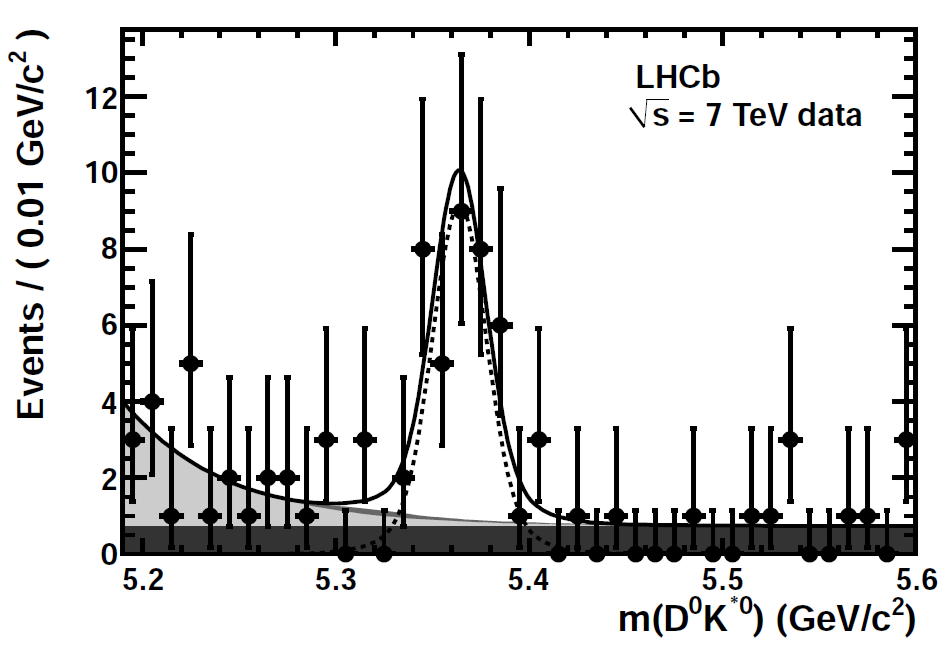}}
\end{center}
\caption{$\overline{B_s}^0 \rightarrow D^0 K^{*0}$ invariant mass distribution.}
\label{DKstarOSplot}       
\end{figure}

The neutral time-integrated mode $\overline{B}^0 \rightarrow D^0 \overline{K}^{*0}$ is also being studied at LHCb using both the ADS and the GLW strategies. As the branching ratio is an order of magnitude smaller than the charged mode, with the data recorded and analysed up to 2011 this mode has not yet been seen. However, the decay $B_s^0 \rightarrow \overline{D}^0 \overline{K}^{*0}$ has been observed and its branching ratio has been measured with respect to the well known $B^0 \rightarrow \overline{D}^0 \rho^0$ decay
\begin{eqnarray}
\frac{\mathcal{B}(\overline{B_s}^0 \rightarrow D^0 K^{*0})}{\mathcal{B}(\overline{B}^0 \rightarrow D^0 \rho^0)} = 1.61 \pm 0.37(\mathrm{stat.}) \\
\pm 0.16(\mathrm{syst.}) \pm 0.20(f_d/f_s) ~\mathrm{,} \nonumber
\end{eqnarray}
see figure \ref{DKstarOSplot}, \textit{c.f.} \cite{DK* paper}. This is the first observation of this mode and it is in agreement with theoretical predictions, showing the good potential of the LHCb experiment for the $\gamma$ searches in a large collection of decay modes. This measurement is important for the $\gamma$ extraction with the neutral mode $\overline{B}^0 \rightarrow D^0 \overline{K}^{*0}$, as it is one of the main backgrounds for this channel.

\section{Conclusion}
\label{concl}
The $\gamma$ angle of the CKM unitary triangle is one of the still least well known parameters of the Standard Model, with an uncertainty in its direct measurements value large compared to that of the other parameters. Efforts are being made at the LHCb experiement to constrain its measurement using several channels, that will give a combined sensitivity considerably better than the current one. The status of these studies using data collected by the experiment at $\sqrt{s}$ = 7 TeV up to 2011 has been presented.

%

%

\end{document}